\documentclass[]{aa}

\usepackage{graphicx,natbib}

\def\jh{\mbox{$\rm (J-H)$}}

\def\jk{\mbox{$\rm (J-K_s)$}}

\def\mv{\mbox{$\rm M_V$}}

\def\rc{\mbox{$\rm R_c$}}
\def\rx{\mbox{$\rm R_{ext}$}}
\def\rhL{\mbox{$\rm R_{hL}$}}
\def\rhSC{\mbox{$\rm R_{hSC}$}}
\def\rhM{\mbox{$\rm R_{hM}$}}

\def\rt{\mbox{$\rm R_t$}}
\def\ms{\mbox{$\rm M_\odot$}}
\def\ds{\mbox{$\rm d_\odot$}}
\def\Rgc{\mbox{$\rm R_{GC}$}}
\def\feh{\mbox{$\rm [Fe/H]$}}
\def\jj{\mbox{$\rm J$}}
\def\hh{\mbox{$\rm H$}}
\def\ks{\mbox{$\rm K_s$}}
\def\dTO{\mbox{$\rm \Delta_{TO}$}}

\begin{document}

\title{Structural parameters of star clusters: relations among light, mass and star-count
radial profiles and the dependence on photometric depth}

\author{C. Bonatto\inst{1} \and E. Bica\inst{1}}
%\offprints{Ch. Bonatto}

\institute{Universidade Federal do Rio Grande do Sul, Departamento de Astronomia\\
CP\,15051, RS, Porto Alegre 91501-970, Brazil\\
\email{charles@if.ufrgs.br, bica@if.ufrgs.br}
\mail{charles@if.ufrgs.br} }

\date{Received --; accepted --}

\abstract
{Structural parameters of model star clusters are measured in radial profiles built from 
number-density, mass-density and surface-brightness distributions, assuming as well different 
photometric conditions.}
{Determine how the core, half-star count and tidal radii, as well as the concentration parameter, 
all of which are derived from number-density profiles, relate to the equivalent radii measured in 
near-infrared surface-brightness and mass-density profiles. We also quantify changes in the resulting
structural parameters due to depth-limited photometry.}
{Star clusters of different ages, structure and mass functions are modelled by assuming that the
radial distribution of stars follows a pre-defined analytical form. Near-infrared surface 
brightness and mass-density profiles result from mass-luminosity relations taken from a set 
of isochrones. Core, tidal and half-light, half-mass and half-star count radii, together with
the concentration parameter, are measured in the three types of profiles, which are built under 
different photometric depths.}
{While surface-brightness profiles are almost insensitive to photometric depth, radii measured
in number-density and mass-density profiles change significantly with it. Compared to radii
derived with deep photometry, shallow profiles result in lower values. This effect increases
for younger ages. Radial profiles of clusters with a spatially-uniform mass function produce
radii that do not depend on depth. With deep photometry, number-density profiles yield radii 
systematically larger than those derived from surface-brightness ones.}
{In general, low-noise surface-brightness profiles result in uniform structural parameters that
are essentially independent of photometric depth. For less-populous star clusters, those projected
against dense fields and/or distant ones, which result in noisy surface-brightness profiles, this 
work provides a quantitative way to estimate the intrinsic radii by means of number-density profiles
built with depth-limited photometry. }


\titlerunning{Structural parameters of model star clusters}

\authorrunning{C. Bonatto \and E. Bica}

\maketitle

\section{Introduction}
\label{intro}

Star clusters are a powerful tool in the investigation of Galaxy structure and dynamics, star 
formation and evolution processes, and as observational constraints to N-body codes. This applies 
especially to the long-lived and populous globular clusters (GCs) that, because of their relatively 
compact nature, can be observed in most regions of the Galaxy, from near the center to the remote 
halo outskirts.

In general terms, the structure of most star clusters can be described by a rather dense core and
a sparse halo, but with a broad range in the concentration level. In this context, the standard 
picture of a GC assumes a isothermal central region and a tidally truncated outer region (e.g. 
\citealt{Binney1998}). Old GCs, in particular, can be virtually considered as dynamically relaxed 
systems (e.g. \citealt{NoGe06}). During their lives clusters are continually affected by internal
processes such as mass loss by stellar evolution, mass segregation and low-mass star evaporation,
and external ones such as tidal stress and dynamical friction e.g. from the Galactic bulge, disk
and giant molecular clouds (e.g. \citealt{Khalisi07}; \citealt{Lamers05}; \citealt{GnOs97}). Over
a Hubble time, these processes tend to decrease cluster mass, which may accelerate the core collapse
phase for some clusters (\citealt{DjMey94}, and references therein). Consequently, these processes,
combined with the presence of a central black hole (in some cases) and physical conditions associated
to the initial collapse, can affect the spatial distribution of light (or mass) both in the central
region and at large radii (e.g. \citealt{GLO99}; \citealt{NoGe06}).

It is clear from the above that crucial information related to the early stages of Galaxy formation,
and to the cluster dynamical evolution, may be imprinted in the present-day internal structure and 
large-scale spatial distribution of GCs (e.g. \citealt{MvdB05}; \citealt{GCProp}). To some extent, 
this reasoning can be extended to the open clusters (OCs), especially the young, which are important 
to determine the spiral arm and disk structures and the rotation curve of the Galaxy (e.g. \citealt{Friel95};
\citealt{DiskProp}). Consequently, the derivation of reliable structural parameters of star clusters, 
GCs in particular, is fundamental to better define their parameter space. This, in turn, may result 
in a deeper understanding of the formation and evolution processes of the star clusters themselves 
and the Galaxy.

Three different approaches have been used to derive structural parameters of star clusters.
The more traditional one is based on the surface-brightness profile (SBP), which considers
the spatial distribution of the brightness of the component stars, usually measured in circular
rings around the cluster center. The compilation of Harris (1996, and the 2003 update\footnote{\em
http://physun.physics.mcmaster.ca/Globular.html}) presents a basically uniform set of parameters for
150 Galactic GCs. Among their structural parameters, the core (\rc), half-light (\rhL) and tidal
(\rt) radii, as well as the concentration parameter $c=\log(\rt/\rc)$, were based mostly on the
SBP database of \citet{TKD95}. SBPs do not necessarily require cluster distances to be known,
since the physically relevant information contained in them is essentially related to the relative
brightness of the member stars. In principle, it is easy to measure integrated light. However, SBPs
are more efficient near the cluster center than in the outer parts, where noise and background
starlight may be a major contributor. Another potential source of noise is the random presence of
bright stars, either from the field or cluster members, especially outside the central region in
the less-populous GCs or most of the OCs. Structural parameters derived from such SBPs would certainly
be affected. One way to minimise this effect is the use of wide rings throughout the whole radius range,
but this would cause spatial resolution degradation on the profiles, especially near the center.

The obvious alternative to SBPs is to use star counts to build  radial density profiles (RDPs), 
in which only the projected number-density of stars is taken into account, regardless of the 
individual star brightness. This technique is particularly appropriate for the outer parts, provided 
a statistically significant, and reasonably uniform, comparison field is available to tackle the 
background contamination. On the other hand, contrary to SBPs, RDPs are less efficient in central 
regions of populous clusters where the density of stars (crowding) may become exceedingly large. 
In such cases it may not be possible to resolve individual stars with the available technology.

Finally, a more physically significant profile can be built by mapping the cluster's stellar mass 
distribution, which essentially determines the gravitational potential and drives most of the dynamical 
evolution. However, mass density profiles (MDPs) not only are affected by the same technical problems 
as the RDPs but, in addition, the cluster distance, age and a reliable mass-luminosity relation are 
necessary to build them.

In principle, the three kinds of profiles are expected to yield different values for the structural 
parameters under similar photometric conditions, since each profile is sensitive to different cluster
parameters, especially the age and dynamical state. Qualitatively, the following effects, basically 
related to dynamical state, can be expected. Large-scale mass segregation drives preferentially low-mass 
stars towards large radii (while evaporation pushes part of these stars beyond the tidal radius, into 
the field), and high-mass stars towards the central parts of clusters. If the stellar mass distribution 
of an evolved cluster can be described by a spatially variable mass function (MF) flatter at the cluster 
center than in the halo, the resulting RDP (and MDP) radii should be larger than SBP ones. The differences
should be more significant for the core than the half and tidal radii, since the core would contain, 
on average, stars more massive than the halo and especially near the tidal radius. Besides, the 
presence of bright stars preferentially in the central parts of young clusters (\citealt{DetAnalOCs}
and references therein) should as well lead to smaller SBP core and half-light radii than the respective 
RDP ones.

Another relevant issue is related to depth-limited photometry. When applied to the observation
of objects at different distances, depth-limited photometry samples stars with different brightness
(or mass), especially at the faint (or low-mass) end. Thus, it would be interesting to quantify the
changes produced in the derived parameters when RDPs, MDPS and SBPs are built with depth-limited
photometry, as well as to check how the structural parameters derived from one type of profile relate
to the equivalent radii measured in the other profiles.

In the present work we face the above issues by deriving structural parameters of star clusters
built under controlled conditions, in which the radial distribution of stars follows a pre-established
analytical profile, and field stars are absent. Effects introduced by mass segregation (simulated
by a spatially variable mass function), age and structure are also considered. This work focuses
on profiles built in the near-infrared range. The main goal of the present work is to examine 
relations among structural parameters measured in the different radial profiles, built under ideal
conditions, especially noise-free photometry and as small as possible statistical uncertainties (using
a large number of stars). In this sense, the results should be taken as upper-limits.

\begin{table*}
\caption[]{Model star cluster specifications}
\label{tab1}
\renewcommand{\tabcolsep}{2.65mm}
\renewcommand{\arraystretch}{1.2}
\begin{tabular}{ccccccccrcccc}
\hline\hline
Model&$R_t/R_c$&c&$\chi_0$&$\chi_t$&Age&\feh&$m_i$&$m_s$&$\langle m\rangle$
&$\rm M_J(TO)$&$\rm M_J(bright)$&$\rm M_J(faint)$\\
   &    &    &     &    &(Myr)&&(\ms)&(\ms)&(\ms)&(mag)&(mag)&(mag)\\
(1)  &(2)  &(3) &(4)  &(5)  &(6)  &(7) &(8)  &(9)  &(10)&(11) &(12) &(13)\\
\hline
GC-A &5&0.7&0.00&1.35&$10^4$&$-1.5$&0.15&1.02&0.43&$+2.86$&$-2.14$&$+9.12$\\
GC-B&20&1.3&0.00&1.35&$10^4$&$-1.5$&0.15&1.02&0.43&$+2.86$&$-2.14$&$+9.12$\\
GC-C&20&1.3&0.00&0.00&$10^4$&$-1.5$&0.15&1.02&0.46&$+2.86$&$-2.14$&$+9.12$\\
GC-D&40&1.6&0.00&1.35&$10^4$&$-1.5$&0.15&1.02&0.43&$+2.86$&$-2.14$&$+9.12$\\
OC-A&15&1.2&0.30&1.35&$10^3$&$~~0.0$&0.15&2.31&0.59&$+0.32$&$-2.68$&$+9.18$\\
OC-B&15&1.2&0.30&1.35&$100$&$~~0.0$&0.15&5.42&0.92&$-1.82$&$-4.82$&$+9.18$\\
OC-C&15&1.2&0.30&1.35&$10$&$~~0.0$&0.15&18.72&1.76&$-4.82$&$-8.82$&$+9.18$\\
\hline
\end{tabular}
\begin{list}{Table Notes.}
\item Col.~3: concentration parameter $c=\log(\rt/\rc)$. Cols.~4 and 5: mass function slopes 
at the cluster center and tidal radius. Cols.~8-10: lower, upper and average star mass. Col.~11:
absolute J magnitude at the turnoff (TO). Cols.~12 and 13: absolute J magnitude at the bright and 
faint ends.
\end{list}
\end{table*}

\begin{figure}
\resizebox{\hsize}{!}{\includegraphics{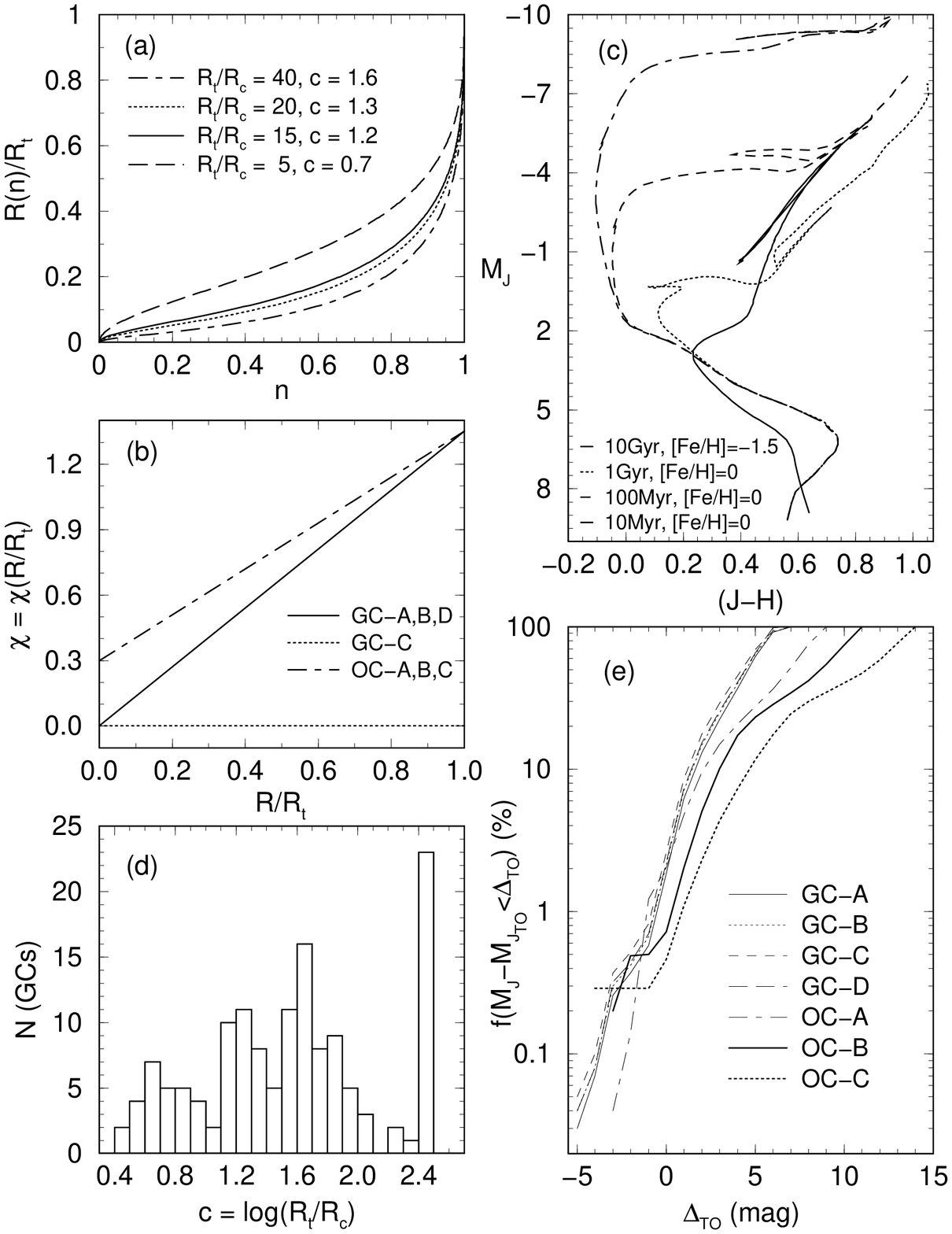}}
\caption{Model star cluster specifications. Panel (a): a random selection of $n$ in the
range $0\leq n\leq1$ produces King-like RDPs in the range $0\leq R\leq\rt$ (see Eq.~\ref{eq2}). 
Panel (b): Radially-variable mass function slopes $\left(\frac{dN}{dm}\propto m^{-(1+\chi)}\right)$
used in the models. Panel (c): Padova isochrones used to simulate the mass-luminosity relation 
of the star cluster models. The 10\,Gyr, $\feh=-1.5$ metallicity isochrone is adopted in the
globular cluster models. Panel (d): distribution of concentration parameters of the GCs in 
H03 with peaks at $c\approx1.6,~1.3,~{\rm and}~0.7$. Panel (e): model fraction of stars 
brighter than $M_J=M_{J_{TO}}+\Delta_{TO}$. In all cases, the fraction of stars brighter
than the TO ($M_{J_{TO}}$) is below the $1\%$ level.}
\label{fig1}
\end{figure}

This work is structured as follows. In Sect.~\ref{ModelSCs} we present the star cluster models
and build radial profiles with depth-limited photometry. In Sect.~\ref{Struc} we derive structural
parameters from each profile, discuss their dependence on depth, and compare similar radii derived
from the different types of profiles. In Sect.~\ref{N6397} we compare relations derived from model 
parameters with those of the nearby GC NGC\,6397. Concluding remarks are given in Sect.~\ref{Conclu}.

\section{The model star clusters}
\label{ModelSCs}

For practical reasons, the model star clusters are simulated by first establishing the number-density
radial distribution. The approach we follow is to build star clusters of different ages and
concentration parameters, with the spatial distribution of stars truncated at the tidal radius 
(\rt). Stars are distributed with distances to the cluster center in the range $0\leq R\leq\rt$, 
with the $R$ coordinate having a number-frequency given by a function similar to a \citet{King62}
three-parameter surface-brightness profile. The mass and brightness of each star are subsequently 
computed according to a pre-defined mass function and mass-luminosity relation consistent with the 
model age. The last step is required for the derivation of the MDP and SBPs.

We point out that different, more sophisticated analytical models have also been used to fit the SBPs
of Galactic and extra-Galactic GCs, other than \citet{King62} profile. The most commonly used are
the single-mass, modified isothermal sphere of \citet{King66} that is the basis of the Galactic
GC parameters given by \citet{TKD95} and H03, the modified isothermal sphere of \citet{Wilson75},
that assumes a pre-defined stellar distribution function (which results in more extended envelopes
than \citealt{King66}), and the power-law with a core of \citet{EFF87} that has been fit to massive
young clusters especially in the Magellanic Clouds (e.g. Mackey \& Gilmore 2003a,b,c). Each function
is characterised by different parameters that are somehow related to the cluster structure. However, 
the purpose here is not to establish a ``best'' fitting function of the structure of star clusters in
general. Instead, we want to quantify changes in the structural parameters, derived from RDPs, MDPs
and SBPs of star clusters with the stellar distribution assumed to follow an analytical function,
under different photometric conditions. We expect that changes in a given parameter should have
a small dependence, if any at all, on the adopted functional form.

The adopted King-like radial distribution function is expressed as
 
\begin{equation}
\label{eq1}
\frac{dN}{2\pi\,R\,dR}=\sigma_0\left[\frac{1}{\sqrt{1+(R/R_c)^2}} -
\frac{1}{\sqrt{1+(R_t/R_c)^2}}\right]^2,
\end{equation}

where $\sigma_0$ is the projected number-density of stars at the cluster center, and \rc\ and
\rt\ are the core and tidal radii, respectively. Since structural differences are basically
controlled by the ratio $\rt/\rc$, we set $\rc=1$ in all models. Such a King-like RDP (for 
$\sigma_0=1.0$) is obtained by numerically inverting the relation (see App.~\ref{Transf})

\begin{equation}
\label{eq2}
n(R) = \frac{x^2-4u(\sqrt{1+x^2}-1)+u^2\ln(1+x^2)}{u^2\ln{u^2}-(u-1)(3u-1)},
\end{equation}

where $x\equiv R/R_c$ and $u^2\equiv 1+(R_t/R_c)^2$. Thus, a random selection of numbers 
in the range $0\leq n\leq1$ produces a King-like radial distribution of stars with the 
radial coordinate in the range $0\leq R/R_t\leq1$. The $R/R_t$ curves as a function of 
$n$ for the models considered in this work are shown in Fig.~\ref{fig1} (Panel a).

Once a given star has been assigned a radial coordinate, its mass is computed with a 
probability proportional to the mass function

\begin{equation}
\label{eq3}
\frac{dN}{dm}\propto m^{-(1+\chi)},
\end{equation}

where the slope varies with $R$ according to $\chi=\chi(R)=\chi_t + (\chi_t-\chi_0)(R/R_t-1)$, 
where $\chi_0$ and $\chi_t$ are the mass function slopes at the cluster center and tidal radius, 
respectively (Table~\ref{tab1} and Fig.~\ref{fig1}).  Thus, the presence of large-scale mass 
segregation in a star cluster can be characterised by a slope $\chi_0$ flatter than $\chi_t$.
Mass values distributed according to Eq.~\ref{eq3} are obtained by randomly selecting numbers in 
the range $0\leq n\leq1$ and using them in the relation of mass with $n~\rm{and~} \chi$ 
(App.~\ref{Transf})

\begin{equation}
\label{eq4}
m=\left\{
\begin{array}{lc}
    m_i\,(m_s/m_i)^n, & \rm{for~\chi=0.0,}\\
    m_s/[(1-n)(m_s/m_i)^\chi+n]^{1/\chi}, & \rm{otherwise},
   \end{array}
   \right .
\end{equation} where $m_i$ and $m_s$ are the lower and upper mass values considered in the models 
(Table~\ref{tab1}).

In what follows we adopt the 2MASS\footnote{\em http://www.ipac.caltech.edu/2mass/releases/allsky/}
photometric system to build SBPs. Finally, the 2MASS \jj, \hh\ and \ks\ magnitudes for each star 
are obtained according to the mass-luminosity relation taken from the corresponding model 
(Table~\ref{tab1}) Padova isochrone (\citealt{Girardi02}). For illustrative purposes the model 
isochrones are displayed in Fig.~\ref{fig1} (panel c).

The set of models considered here is intended to be objectively representative of the star cluster 
parameter space. For globular clusters we use the standard age of 10\,Gyr and  the spatially 
uniform metallicity $\feh=-1.5$, which is typical of the metal-poor Galactic GCs (e.g. 
\citealt{GCProp}). However, we note that abundance variations have been suggested to occur within 
GCs (e.g. \citealt{Gratton04}). Basically, small to moderate metallicity gradients would produce slight 
changes in the colour and magnitude of the stars in different parts of the cluster, which has no effect 
on the (star-count derived) RDPs and MDPs. The effect on the SBPs may be small as well, provided that 
the magnitude bin used to build the SBPs is wide enough to accommodate such magnitude changes. 
As for the core/tidal structure 
we consider the ratios $R_t/R_c=40,~20,~15,~{\rm and}~5$, or equivalently the concentration parameters
$c=\log{(R_t/R_c)\approx1.6,~1.3,~1.2,~{\rm and}~0.7}$, which roughly correspond to the peaks in the 
distribution of $c$ values presented by the regular (non-post core collapse) GCs given in H03 
(Fig.~\ref{fig1}, panel d). Models GC-A, B and D take into account mass segregation by means of a 
flat ($\chi_0=0.00$) mass function at the center and a Salpeter (1955) IMF ($\chi_t=1.35$) at the 
tidal radius. GC-C model is similar to GC-B, except that it considers a uniform, heavily depleted 
MF ($\chi_0=0.00$) throughout the cluster. OCs are represented by solar-metallicity models with the
ages 10\,Myr (to allow for the presence of bright stars in young OCs), 100\,Myr (somewhat evolved 
OCs) and 1\,Gyr (intermediate-age OCs), $R_t/R_c=15$ ($c\approx1.2$) and a spatially variable MF
(Table~\ref{tab1}). The values of $c$ and the core/halo MF slopes are representative of OCs
(\citealt{DetAnalOCs}).  Another effect not considered here is differential absorption. 
In principle, low to moderate differential absorption should have a minimum effect on the radial 
profiles, because of the same reasons as those given above for the metallicity gradient. High values,
on the other hand, would affect RDPs as well, because of a radially-dependent loss of stars due to 
depth-limited photometry. However, inclusion of this effect is beyond the scope of the present work. 

As expected, the fraction of stars brighter than the turnoff (TO) in the resulting star cluster 
models is significantly smaller than 1\% (Fig.~\ref{fig1}, panel e). Thus, we had to use a total 
number of stars of $1\times10^9$ in all models, so that the radial profiles resulted statistically
significant (small $1\sigma$ Poisson error bars) especially at the shallowest magnitude depth.

\subsection{Depth-varying radial profiles}
\label{DeptVP}

The radial profiles were built considering all stars brighter than a given magnitude threshold, 
with the TO as reference. At the bright end, statistically significant GC profiles were obtained 
for $\dTO\equiv M_{J,th}-M_{J,TO}=-5$, where $M_{J,th}$ and $M_{J,TO}$ are the threshold 
and TO absolute magnitudes in the 2MASS \jj\ band. At the faint end, GC-models have $\dTO=6.3$. 
OC models have $\dTO=-3~{\rm and~} -4$ at the bright end, and $\dTO=8.9,~11.0,~{\rm and}~14.0$, 
at the faint end. 

Starting at the bright magnitude end, RDPs, MDPs and SBPs were built considering stars with
the \jj\ magnitude brighter than a given faint threshold, with the magnitude depth increasing 
in steps of $\dTO=1$, up to the respective faint magnitude end. 

\begin{figure}
\resizebox{\hsize}{!}{\includegraphics{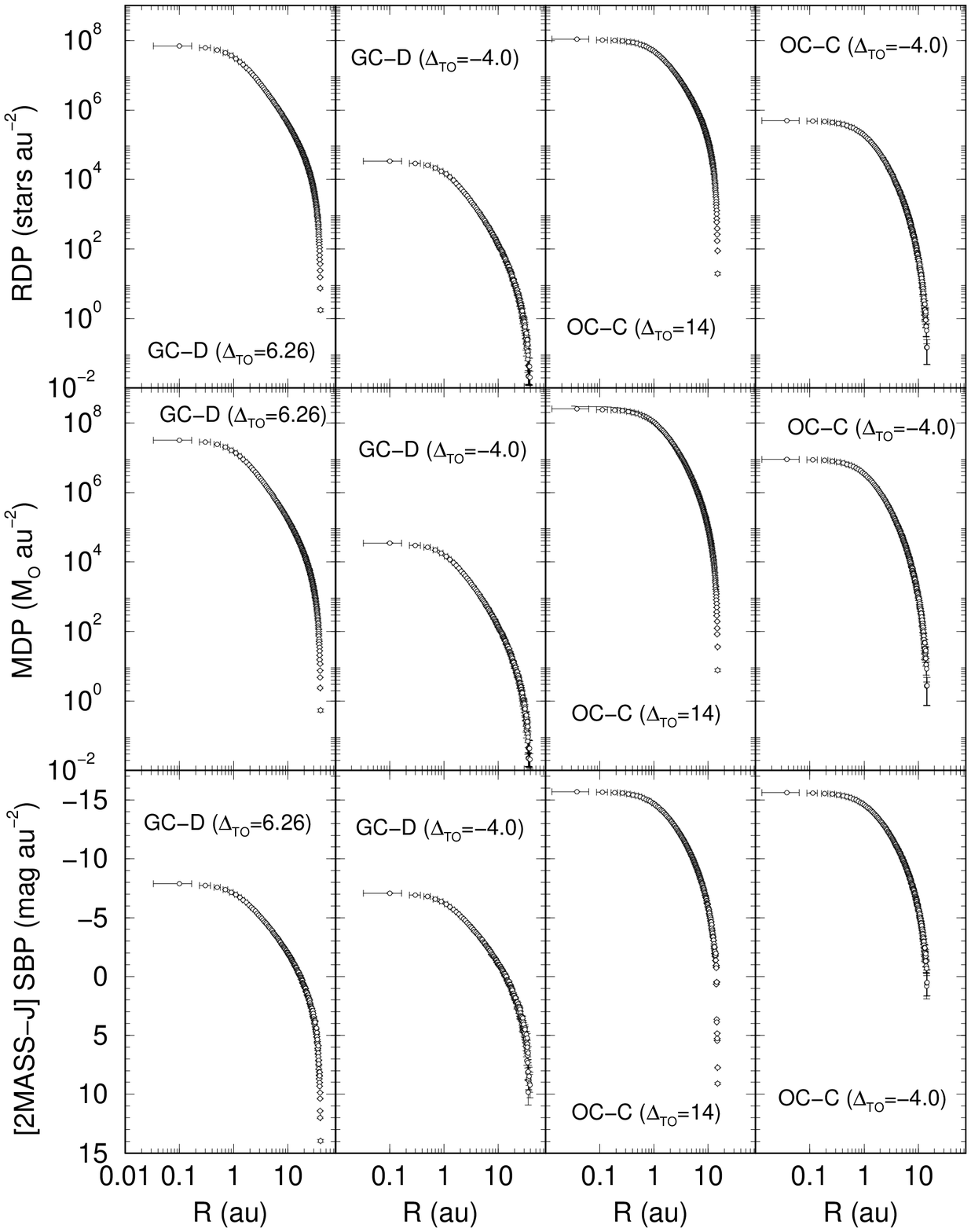}}
\caption{A selection of RDPs (top panels), MDPs (middle) and 2MASS \jj\ magnitude SBPs (bottom) 
that illustrate structural changes under different magnitude depths. Arbitrary units (au) are 
used both for the radial coordinate and projected area.}
\label{fig2}
\end{figure}

Figure~\ref{fig2} displays a selection of profiles corresponding to both extremes in magnitude
depths, for the GC-D and OC-C models. These profiles are representative of the whole set of models, 
especially in terms of the small uncertainties associated with each radial coordinate. Reflecting
the large differences in the number of stars at different photometric depths, the central values
of the number and mass densities, and surface-brightness, vary significantly from the shallowest
to the deepest profiles.

\begin{table*}
\caption[]{Model star cluster structural parameters for different photometric depths}
\label{tab2}
\renewcommand{\tabcolsep}{1.3mm}
\renewcommand{\arraystretch}{1.2}
\begin{tabular}{cccccccccccc}
\hline\hline
&\multicolumn{3}{c}{RDP}&&\multicolumn{3}{c}{MDP}&&\multicolumn{3}{c}{SBP (\jj\ band)}\\
\cline{2-4}\cline{6-8}\cline{10-12}
$\Delta_{TO}$&\rc&\rhSC&\rt&&\rc&\rhM&\rt&&\rc&\rhL&\rt\\
(mag)&(au)&(au)&(au)&&(au)&(au)&(au)&&(au)&(au)&(au)\\
(1)  &(2) &(3) &(4) &&(5) &(6) &(7) &&(8) &(9) &(10) \\
\hline
&\multicolumn{11}{c}{Model: GC-A; Input RDP parameters:  $\rc=1.0$, $\rt=5.0$}\\
\cline{2-12}
$-5.0$&$0.78\pm(\dag)$&$1.02\pm(\dag)$&$4.61\pm0.01$&&$0.78\pm(\dag)$&$1.02\pm(\dag)$&$4.61\pm0.01$&&$0.75\pm0.01$&$1.01\pm(\dag)$&$4.80\pm0.01$\\
$~0.0$&$0.76\pm(\dag)$&$1.02\pm(\dag)$&$4.77\pm0.01$&&$0.76\pm(\dag)$&$1.02\pm(\dag)$&$4.77\pm0.01$&&$0.75\pm0.01$&$1.01\pm(\dag)$&$4.79\pm0.01$\\
$+6.3$&$1.00\pm(\dag)$&$1.19\pm(\dag)$&$5.00\pm0.01$&&$0.92\pm(\dag)$&$1.14\pm(\dag)$&$4.91\pm(\dag)$&&$0.75\pm0.01$&$1.03\pm(\dag)$&$4.80\pm0.01$\\
\hline
&\multicolumn{11}{c}{Model: GC-B; Input RDP parameters:  $\rc=1.0$, $\rt=20.0$}\\
\cline{2-12}
$-5.0$&$0.87\pm0.01$&$2.03\pm(\dag)$&$17.31\pm0.08$&&$0.87\pm0.01$&$2.03\pm(\dag)$&$17.31\pm0.08$&&$0.86\pm0.01$&$2.04\pm0.01$&$17.82\pm0.05$\\
$~0.0$&$0.83\pm0.01$&$2.03\pm(\dag)$&$18.72\pm0.08$&&$0.83\pm0.01$&$2.03\pm(\dag)$&$18.72\pm0.08$&&$0.86\pm0.01$&$2.03\pm(\dag)$&$17.80\pm0.03$\\
$+6.3$&$1.00\pm(\dag)$&$2.39\pm(\dag)$&$20.00\pm(\dag)$&&$0.95\pm(\dag)$&$2.27\pm(\dag)$&$19.28\pm0.03$&&$0.86\pm0.01$&$2.05\pm(\dag)$&$17.80\pm0.02$\\
\hline
&\multicolumn{11}{c}{Model: GC-C; Input RDP parameters:  $\rc=1.0$, $\rt=20.0$}\\
\cline{2-12}
$-5.0$&$1.00\pm(\dag)$&$2.38\pm0.01$&$20.02\pm0.03$&&$1.00\pm(\dag)$&$2.38\pm0.01$&$20.02\pm0.04$&&$1.00\pm(\dag)$&$2.38\pm0.01$&$19.94\pm0.06$\\
$~0.0$&$1.00\pm(\dag)$&$2.39\pm(\dag)$&$20.00\pm0.01$&&$1.00\pm(\dag)$&$2.39\pm(\dag)$&$20.00\pm0.01$&&$1.00\pm(\dag)$&$2.39\pm(\dag)$&$19.95\pm0.03$\\
$+6.3$&$1.00\pm(\dag)$&$2.39\pm(\dag)$&$20.00\pm(\dag)$&&$1.00\pm(\dag)$&$2.39\pm(\dag)$&$20.00\pm(\dag)$&&$1.00\pm(\dag)$&$2.39\pm(\dag)$&$19.97\pm0.03$\\
\hline
&\multicolumn{11}{c}{Model: GC-D; Input RDP parameters:  $\rc=1.0$, $\rt=40.0$}\\
\cline{2-12}
$-5.0$&$0.90\pm0.01$&$2.81\pm0.02$&$33.96\pm0.19$&&$0.90\pm0.01$&$2.81\pm0.02$&$33.96\pm0.19$&&$0.91\pm0.01$&$2.82\pm(\dag)$&$34.18\pm0.05$\\
$~0.0$&$0.86\pm0.01$&$2.82\pm(\dag)$&$37.15\pm0.17$&&$0.86\pm0.01$&$2.82\pm(\dag)$&$37.15\pm0.17$&&$0.91\pm0.01$&$2.82\pm(\dag)$&$34.00\pm0.05$\\
$+6.3$&$1.00\pm(\dag)$&$3.30\pm(\dag)$&$39.99\pm0.01$&&$0.96\pm(\dag)$&$3.14\pm(\dag)$&$38.51\pm0.07$&&$0.91\pm0.01$&$2.82\pm(\dag)$&$34.20\pm0.04$\\
\hline
&\multicolumn{11}{c}{Model: OC-A; Input RDP parameters: $\rc=1.0$, $\rt=15.0$}\\
\cline{2-12}
$-3.0$&$0.82\pm0.01$&$1.70\pm(\dag)$&$12.85\pm0.07$&&$0.82\pm0.01$&$1.70\pm(\dag)$&$12.85\pm0.07$&&$0.81\pm0.01$&$1.72\pm(\dag)$&$13.18\pm0.02$\\
$~0.0$&$0.78\pm0.01$&$1.72\pm(\dag)$&$13.78\pm0.06$&&$0.78\pm0.01$&$1.72\pm(\dag)$&$13.78\pm0.06$&&$0.82\pm0.01$&$1.72\pm(\dag)$&$13.19\pm0.02$\\
$+8.9$&$1.00\pm(\dag)$&$2.08\pm(\dag)$&$15.00\pm0.01$&&$0.91\pm(\dag)$&$1.93\pm(\dag)$&$14.43\pm0.03$&&$0.81\pm0.01$&$1.73\pm(\dag)$&$13.20\pm0.01$\\
\hline
&\multicolumn{11}{c}{Model: OC-B; Input RDP parameters:  $\rc=1.0$, $\rt=15.0$}\\
\cline{2-12}
$-3.0$&$0.72\pm0.01$&$1.61\pm(\dag)$&$13.30\pm0.08$&&$0.72\pm0.01$&$1.61\pm(\dag)$&$13.30\pm0.08$&&$0.76\pm0.01$&$1.61\pm(\dag)$&$12.75\pm0.03$\\
$~0.0$&$0.70\pm0.02$&$1.61\pm(\dag)$&$13.67\pm0.08$&&$0.70\pm0.02$&$1.61\pm(\dag)$&$13.67\pm0.08$&&$0.77\pm0.01$&$1.61\pm(\dag)$&$12.74\pm0.03$\\
$+11.0$&$1.00\pm(\dag)$&$2.08\pm(\dag)$&$15.00\pm(\dag)$&&$0.84\pm0.01$&$1.84\pm(\dag)$&$14.33\pm0.04$&&$0.74\pm0.02$&$1.63\pm(\dag)$&$12.75\pm0.02$\\
\hline
&\multicolumn{11}{c}{Model: OC-C; Input RDP parameters:  $\rc=1.0$, $\rt=15.0$}\\
\cline{2-12}
$-4.0$&$0.62\pm0.02$&$1.49\pm(\dag)$&$13.06\pm0.10$&&$0.62\pm0.02$&$1.49\pm(\dag)$&$13.05\pm0.10$&&$0.71\pm0.01$&$1.49\pm(\dag)$&$11.99\pm0.03$\\
$~0.0$&$0.62\pm0.02$&$1.49\pm(\dag)$&$13.10\pm0.10$&&$0.62\pm0.02$&$1.49\pm(\dag)$&$13.09\pm0.10$&&$0.71\pm0.01$&$1.49\pm(\dag)$&$11.99\pm0.03$\\
$+14.0$&$1.00\pm(\dag)$&$2.08\pm(\dag)$&$15.00\pm(\dag)$&&$0.70\pm0.02$&$1.70\pm(\dag)$&$14.35\pm0.05$&&$0.64\pm0.03$&$1.49\pm(\dag)$&$12.00\pm0.02$\\
\hline
\end{tabular}
\begin{list}{Table Notes.}
\item ($\dag$): uncertainty smaller than 0.01 arbitrary units (au). The half-type radii are half-star 
counts (\rhSC), half-mass (\rhM) and half-light (\rhL).
\end{list}
\end{table*}

\begin{figure}
\resizebox{\hsize}{!}{\includegraphics{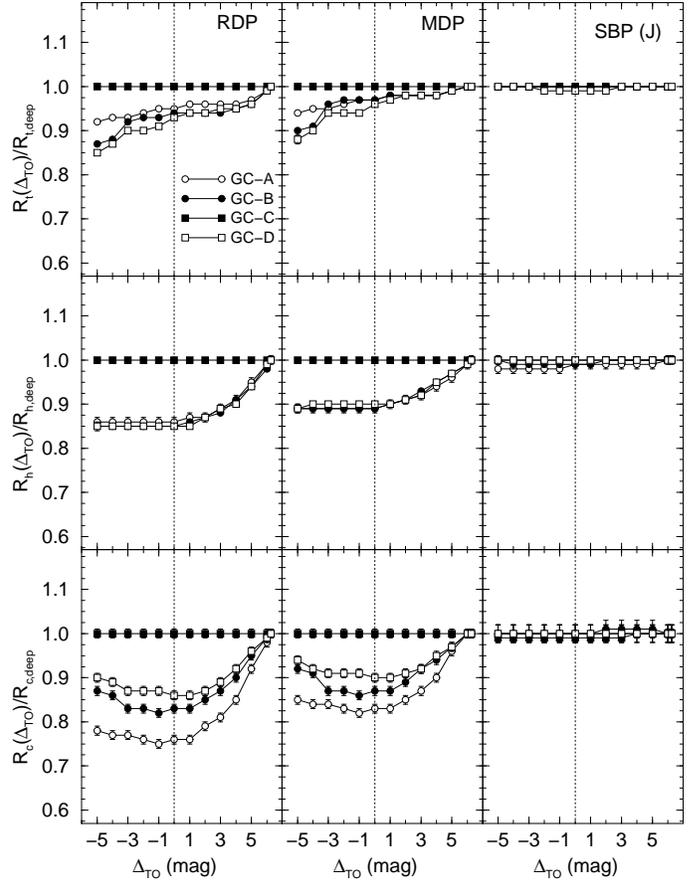}}
\caption{Structural parameters of the GC models. Top panels: Ratio of the tidal radius measured
in profiles with a photometric depth \dTO\ with respect to that derived from the deepest one, for
the RDPs (left panels), MDPs (vertical-middle) and SBPs (right). Horizontal-middle panels: half-type
radii. Bottom: core radii. TO values are indicated by the dotted line. Except for GC-C (uniform
mass function), the remaining models present changes in radii in the RDPs and MDPs. SBP radii
are essentially uniform.}
\label{fig3}
\end{figure}

\begin{figure}
\resizebox{\hsize}{!}{\includegraphics{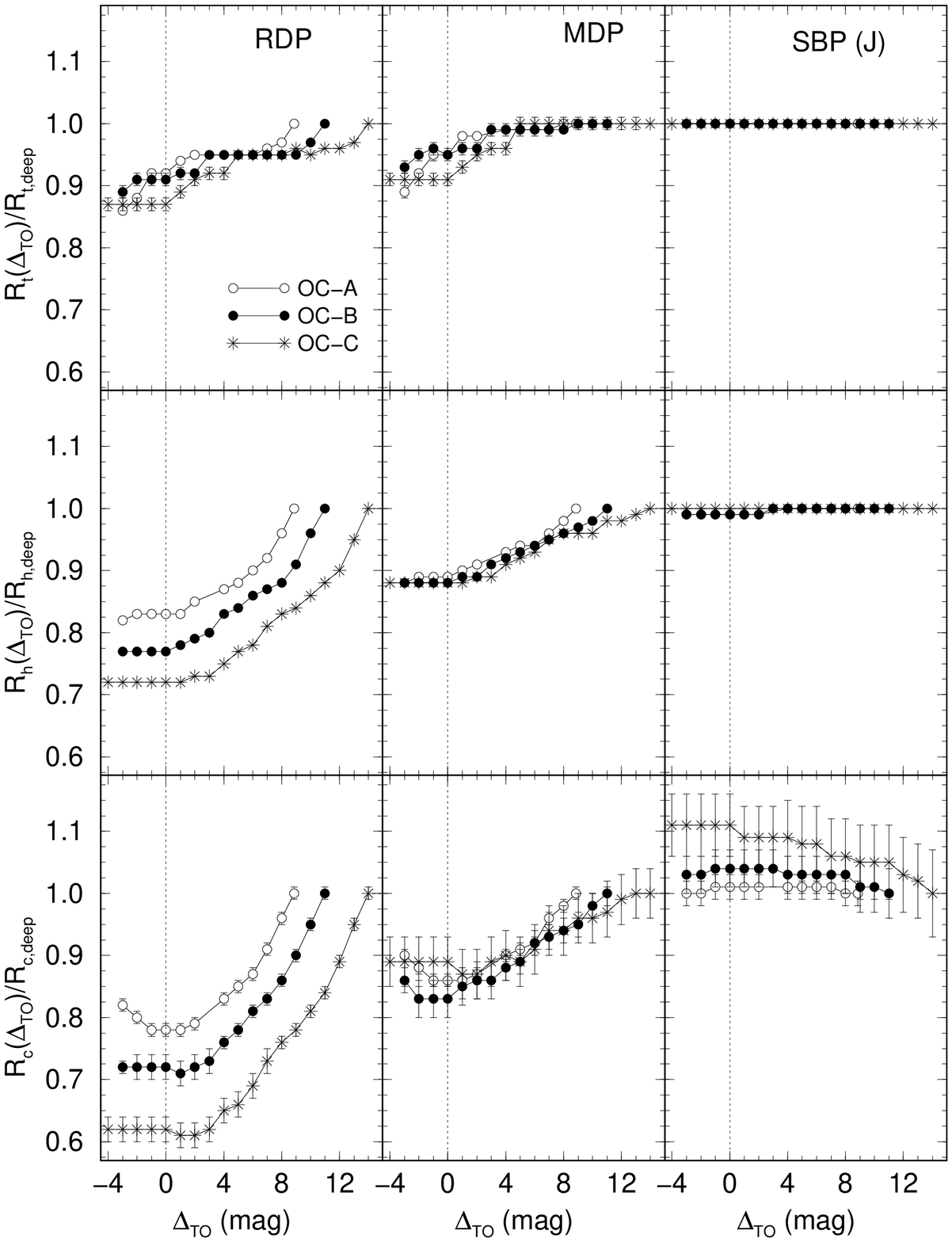}}
\caption{Same as Fig.~\ref{fig3} for the OC models. For comparison purposes, the y-scale
is the same as in Fig.~\ref{fig3}. Similarly to the GC models, radii changes are conspicuous
in the RDPs and MDPs. }
\label{fig4}
\end{figure}

\section{Structural parameters {\em vs.} photometry depth}
\label{Struc}

The depth-varying model SBPs are fit with the empirical three-parameter function introduced by 
\cite{King62} to describe the surface-brightness distribution of GCs, which is characterised 
by the presence of the core and tidal radii. For RDPs and MDPs we use the King-like analytical
profile that describes the projected number-density of stars as a function of \rc\ and \rt,
$\sigma(R)=\frac{dN}{2\pi\,R\,dR}$, as given by eq.~\ref{eq1}. We also compute the distances from
the center which contains half of the cluster's total light, stars and mass. The half-star count
(\rhSC), light (\rhL) and mass (\rhM) radii are derived by directly integrating the corresponding
profiles.

%The SBPs are fitted  with the empirical King profile (King 1962) which is different from the
%dynamical King models (1966), however xxx (Cohen)
%$\mu(R)=\mu_0-5\log{((1+(R/R_c)^2)^{-1/2} - (1+(R_t/R_c)^2)^{-1/2})}$

\begin{figure}
\resizebox{\hsize}{!}{\includegraphics{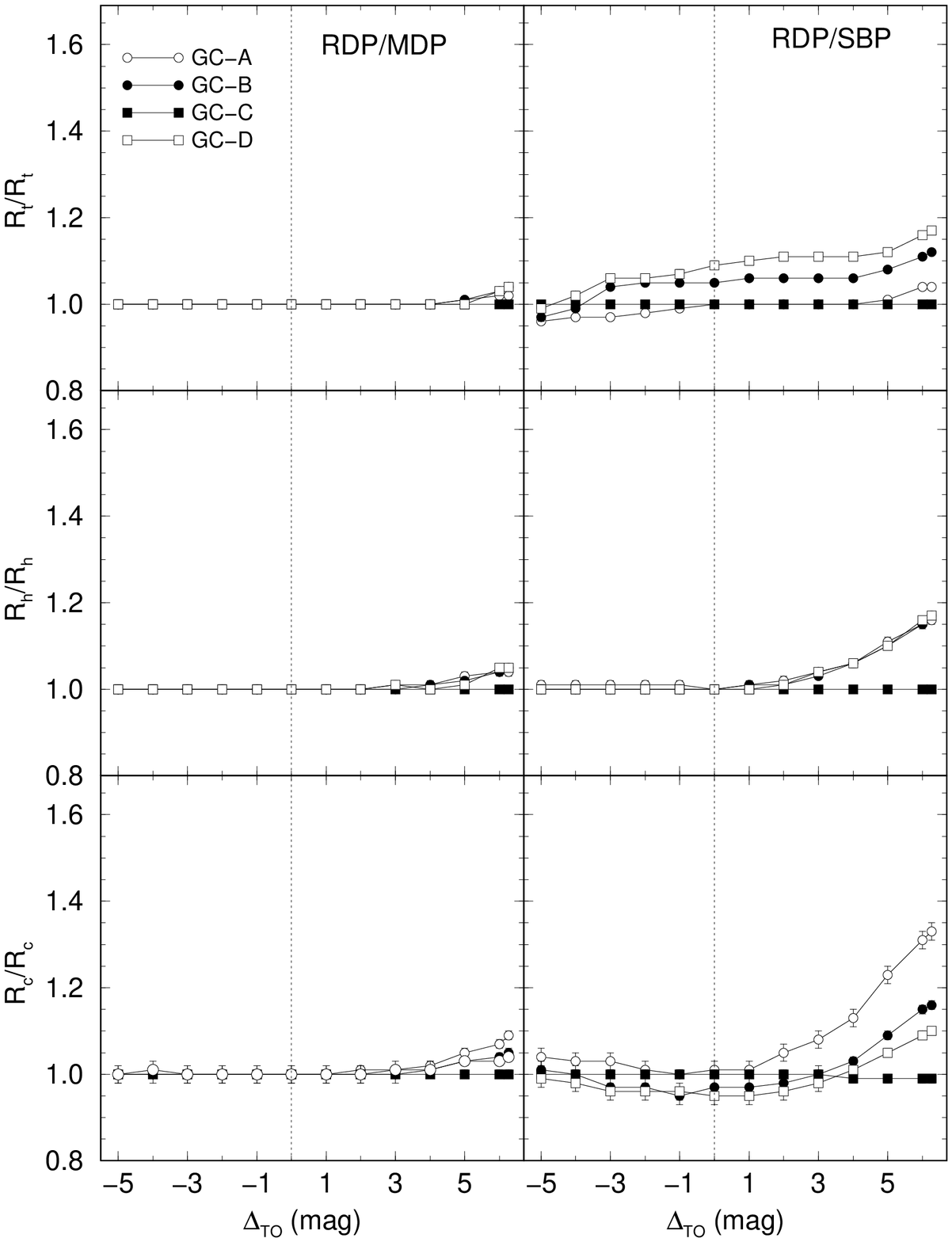}}
\caption{GC model profiles. Ratio between the same type of radii as measured in RDPs and MDPs
(left panels) and RDPs and SBPs (right panels). From top to bottom: tidal, half and core radii.
TO values are indicated by the dotted line.}
\label{fig5}
\end{figure}

\begin{figure}
\resizebox{\hsize}{!}{\includegraphics{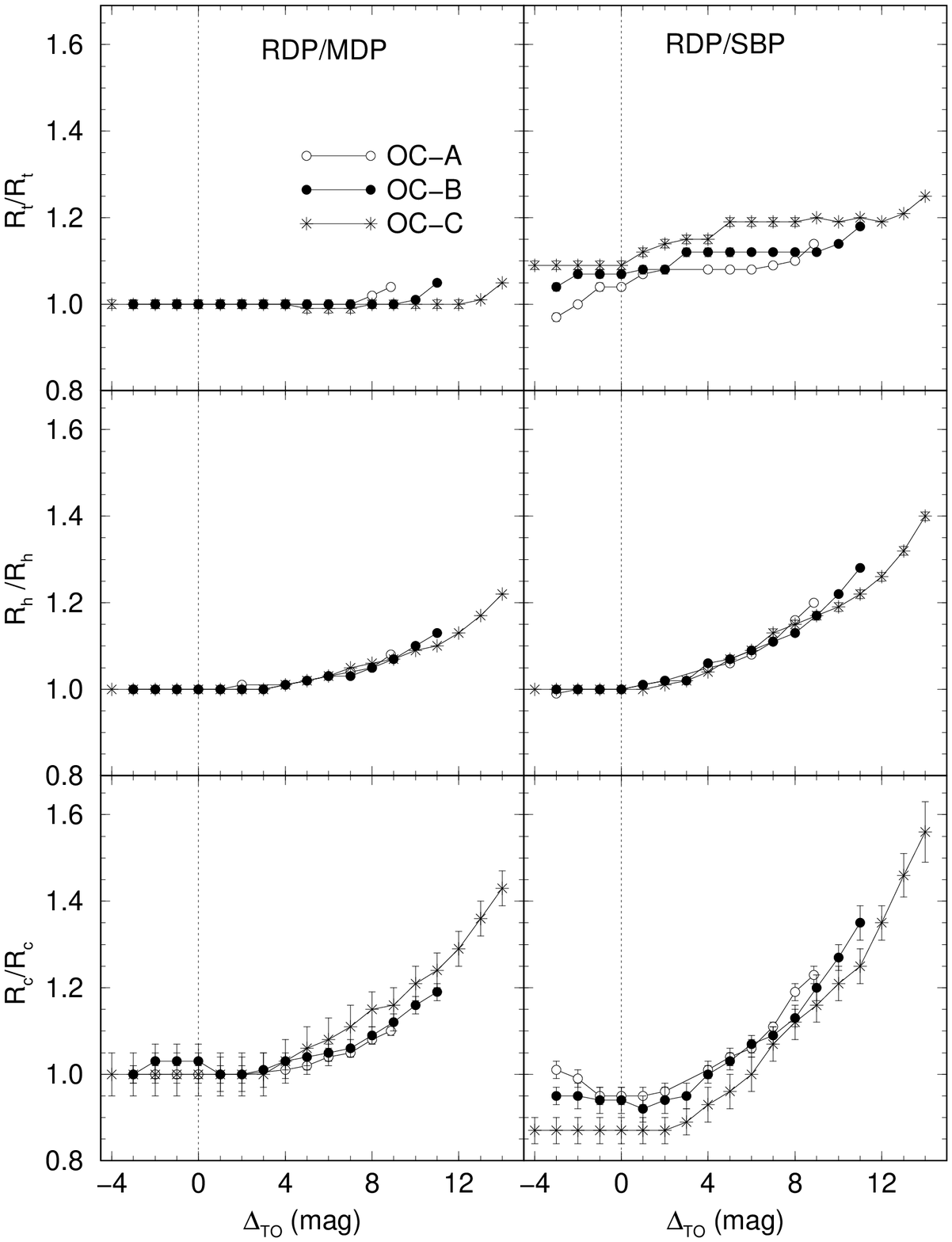}}
\caption{Same as Fig.~\ref{fig5} for the OC models. For comparison purposes, the y-scale
is the same as in Fig.~\ref{fig5}.}
\label{fig6}
\end{figure}

A selection of the resulting structural parameters as a function of \dTO\ is given in 
Table~\ref{tab2}. For simplicity we only present the values obtained from the bright and faint 
magnitude ranges, as well as for $M_J\leq M_{J,TO}$. The whole set of parameters are contained
in Figs.~\ref{fig3} - \ref{fig6}. At first glance, RDP and MDP radii present a significant 
decrease for shallower photometry, with respect to the intrinsic values. SBP radii, on the other 
hand, are more uniform. The most noticeable feature is that, except for GC-C (uniform mass function), 
RDP and MDP radii tend to become increasingly larger than SBP ones with increasing photometric
depth.

\subsection{Dependence on photometric depth}
\label{DependDepth}

In Fig.~\ref{fig3} we compare the radii measured in GC profiles built with a given photometric
depth (e.g. $\rc(\Delta_{TO})$) with the intrinsic ones, i.e. those derived from the deepest 
profiles ($R_{c,deep}$). RDP parameters are more affected than the MDP ones, while the SBP ones 
are essentially uniform, thus insensitive to photometric depth. Among the radii, RDP and MDP 
core are the most affected (underestimated), followed by the half and tidal radii. In the most 
concentrated model (GC-A, $c\approx0.7$), measurements or \rc\ in the RDP may be underestimated 
by a factor $\approx25\%$ in profiles shallower than near the TO, with respect to $R_{c,deep}$, 
and $\approx20\%$ in MDPs. The effect is smaller in \rhSC\ and \rhM, which may be underestimated 
by $\approx15\%$ in the same profiles. The underestimation in the tidal radii is smaller than
$\approx10\%$. As expected, RDP, MDP and SBP radii do not change when the mass function is 
uniform (GC-C model). 

Similar radii ratios in the OC models are examined in Fig.~\ref{fig4}. Qualitatively, the
same conclusions drawn from the GC models apply to the OC ones. However, the underestimation
factor of RDP radii increases for younger ages, to the point that \rc\ drops to $\approx60\%$ 
of the deepest value for all profiles shallower than $\approx3$\, mag below the TO in the OC-C 
model ($10$\,Myr), and to $\approx70\%$ for OC-B ($100$\,Myr). The respective half-star count 
radii are affected by similar, although smaller, underestimation factors. MDP radii are less 
affected by cluster age than RDP ones. Similarly to the GC models (Fig.~\ref{fig3}), the three 
types of SBP radii are essentially insensitive to photometric depth, within uncertainties. We 
note that the presence of bright stars in the central region of young clusters (OC-C) appears 
to introduce a small dependence of the core radius on  photometric depth (bottom-right panel).

\begin{figure}
\resizebox{\hsize}{!}{\includegraphics{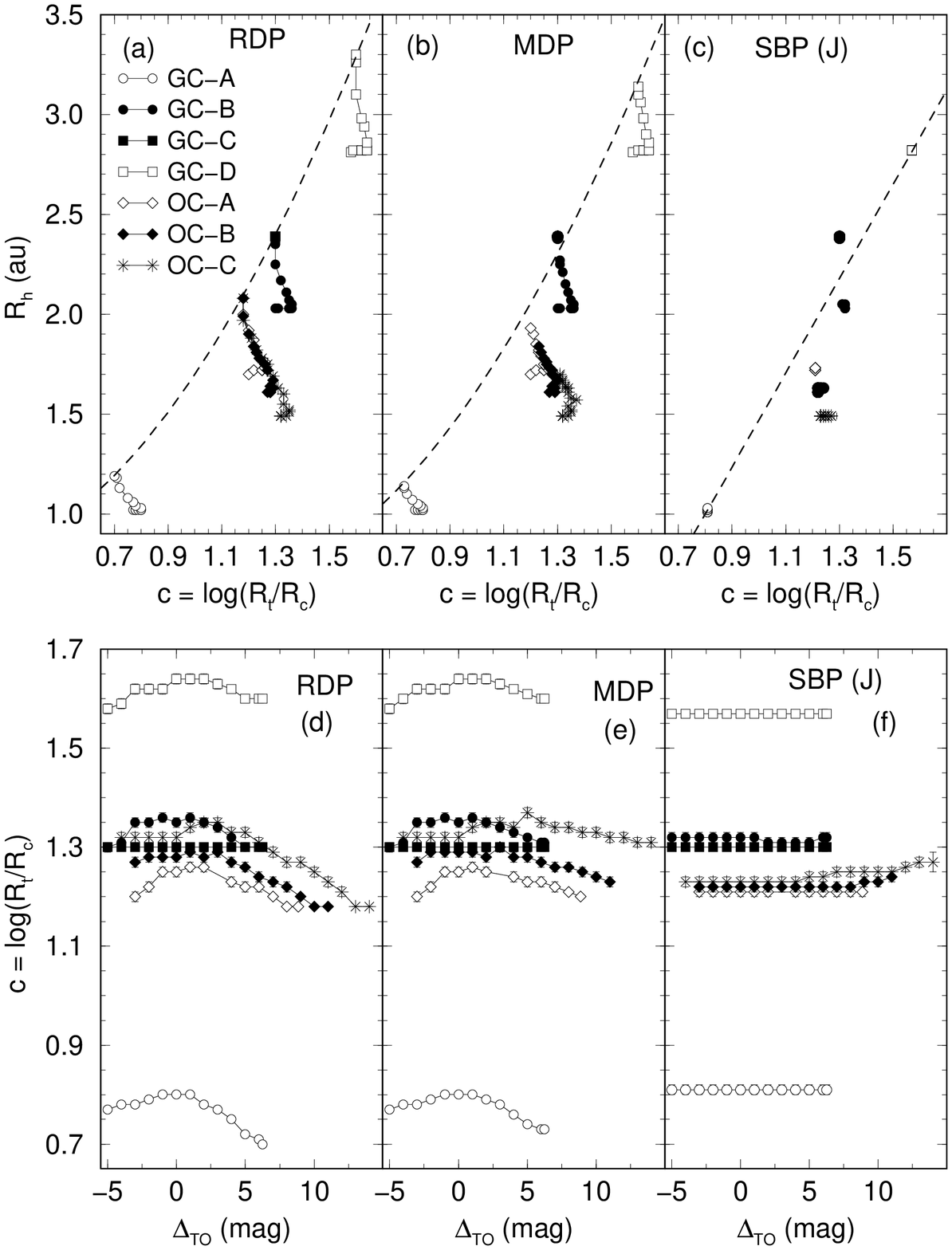}}
\caption{Top panels: relation of the half-type radii with the concentration parameter,
for the RDPs (left panel), MDPs (middle) and SBPs (right). For each model, $R_h$ values 
increase for deeper profiles. Dashed line in panels (a) and (b): $R_h\sim c^2$. In
panel (c): $\rhL\sim c$. Bottom panels: concentration parameter as a function of photometric 
depth.}
\label{fig7}
\end{figure}

\subsection{Comparison of similar radii among different profiles}
\label{CompDifProf}
 
Differences on the same type of radii among the profiles, introduced essentially by a
spatially variable MF, are discussed in Fig.~\ref{fig5} for the GC models. Regardless 
of the model assumptions, RDP and MDP radii are essentially the same, except for the 
profiles corresponding to deep photometry, for which the RDP radii become slightly larger 
than the MDP ones. This occurs basically because of the larger fraction of low-mass stars 
at the outer parts of the clusters. Since all stars have equal weight in the building of 
the RDPs, the accumulation of low-mass stars at large radii ends up broadening the RDPs 
with respect to the MDPs. On the other hand, RDP core and half-star count radii tend to 
be larger than the SBP ones for profiles including stars fainter than near the TO. RDP 
\rt\ may be 10 -- 20\% larger than SBP ones for all depths. As discussed above, the 
uniformly-depleted MF of GC-C model produces profiles whose radii are independent of 
photometry depth. The RDP to SBP core and tidal radii ratios decrease with concentration
parameter. The RDP to SBP half-type radii ratios do not depend on $c$. 

The same analysis applied to the OC models is discussed in Fig.~\ref{fig6}. The presence
of massive stars in young clusters enhances the RDP to MDP radii ratios, especially the
core and to some extent, the half-type radii. This occurs for profiles that contain stars 
brighter than $\approx4$\,mag below the TO. For the youngest model (OC-C), the core radius
measured in the RDP may be $\approx40\%$ larger than the MDP one. This effect is enhanced 
when RDP radii are compared to SBP ones, again decreasing in intensity from the core to 
tidal radii. For OC-C, RDP core, half and tidal radii are $\approx55\%$, $\approx40\%$, 
and $\approx25\%$ larger than the equivalent SBP ones. Comparing with the GC models 
(Fig.~\ref{fig5}), the presence of a larger fraction of more massive (brighter) stars
towards the center in young clusters tend to enhance radii ratios of RDP with respect to
MDP, and especially, RDP to SBP.

\subsection{Further relations}
\label{FurtRel}

The models discussed in previous sections can be used as well to examine the dependence of
the half-type radii with the concentration parameter, and to test how $c$ varies with  
photometric depth. These issues are presented in Fig.~\ref{fig7}.

As already suggested by Figs.~\ref{fig3} and \ref{fig4}, the relation of the half radius  
with $c$, in a given model, changes significantly with photometric depth in RDPs (panel a)
and MDPs (panel b). In SBPs, on the other hand, it is almost insensitive to depth (panel c). 
From eq.~\ref{eq1}, the half-star count radius is tightly related to the concentration
parameter according to $\rhSC=(0.69\pm0.01)+(1.01\pm0.01)\,c^2$. This curve fits
well the values measured in the deepest RDP of all GC and OC models alike (panel a). Such
a relation fails for the shallower profiles. A similar, but poorer, relation applies to the
values derived from the deepest MDPs (panel b), $\rhM=(0.63\pm0.09)+(0.99\pm0.05)\,c^2$.
It fails especially for the young (OC) models. The GC SBPs, on the other hand, can be poorly
fit with the linear function $\rhL=(-0.9\pm0.1)+(2.4\pm0.1)\,c$ (panel c).

Concentration parameters measured in RDPs and MDPs (panels d and e) change with photometric depth.
Around the TO they reach the maximum value, which corresponds to a star cluster $\approx15\%$ more 
concentrated than the pre-established value (Table~\ref{tab1}). At the shallowest profiles $c$
presents a value intermediate between the maximum and the pre-established one, which is retrieved 
at the deepest profiles with the inclusion of the numerous low-mass stars. The exception again is 
the uniform MF model GC-C, whose $c$ values do not change with $\Delta_{TO}$. $c$ values measured
in SBPs are essentially insensitive to photometric depth (panel f).

\section{NGC\,6397: a test case}
\label{N6397}

We compare the results derived for the model star clusters with similar parameters measured in 
the $\mv=-6.63$, nearby GC ($\ds=2.3$\,kpc) NGC\,6397. Being populous is important to produce 
statistically significant radial profiles, while the proximity allows a few magnitudes fainter 
than the giant branch to be reached with depth-limited photometry.

NGC\,6397 is a post-core collapse GC with evidence of large-scale mass segregation, as indicated
by a mass function flatter at the center than outwards (\citealt{Andreuzzi04} and references 
therein). 

Additional relevant data (from H03) for the metal-poor ($\feh=-1.95$) GC NGC\,6397 are the 
Galactocentric distance $\Rgc=6$\,kpc, half-light and tidal radii (measured in the V band) 
$\rhL=2.33\arcmin$ and $\rt=15.81\arcmin$, and Galactic coordinates $\ell=338.17^\circ$, 
$b=-11.96^\circ$. Thus, bulge star contamination is not heavy, and cluster sequences can be 
unambiguously detected, which is important for the extraction of radial profiles with small 
errors (see below). Using SBPs built with 2MASS images and a fit with \citet{King62} profile, 
\citet{Cohen07} derived the core radius in the \jj\ band $\rc(J)=61.5\arcsec\pm9.3\arcsec$. 
However, based on Hubble Space Telescope data and using a power-law plus core as fit function, 
\citet{NoGe06} derived $\rc=3.7\arcsec$ in the equivalent V band, thus roughly resolving the 
post-core collapse nucleus.

The post-core collapse state of NGC\,6397 does not affect the present analysis, since the goal here 
is the determination of changes produced in cluster radii derived under the assumption of a King-like 
profile (Sect.~\ref{Struc}) applied to RDP, MDP and SBPs built with different magnitude depths. We base 
the analysis of NGC\,6397 on \jj, \hh\ and \ks\ 2MASS photometry extracted using VizieR\footnote{\em
vizier.u-strasbg.fr/viz-bin/VizieR?-source=II/246} in a circular field of radius $\rx=70\arcmin$
centered on the coordinates provided in H03. This extraction radius is large enough to encompass
the whole cluster, allowing as well for a significant comparison field.

\begin{figure}
\resizebox{\hsize}{!}{\includegraphics{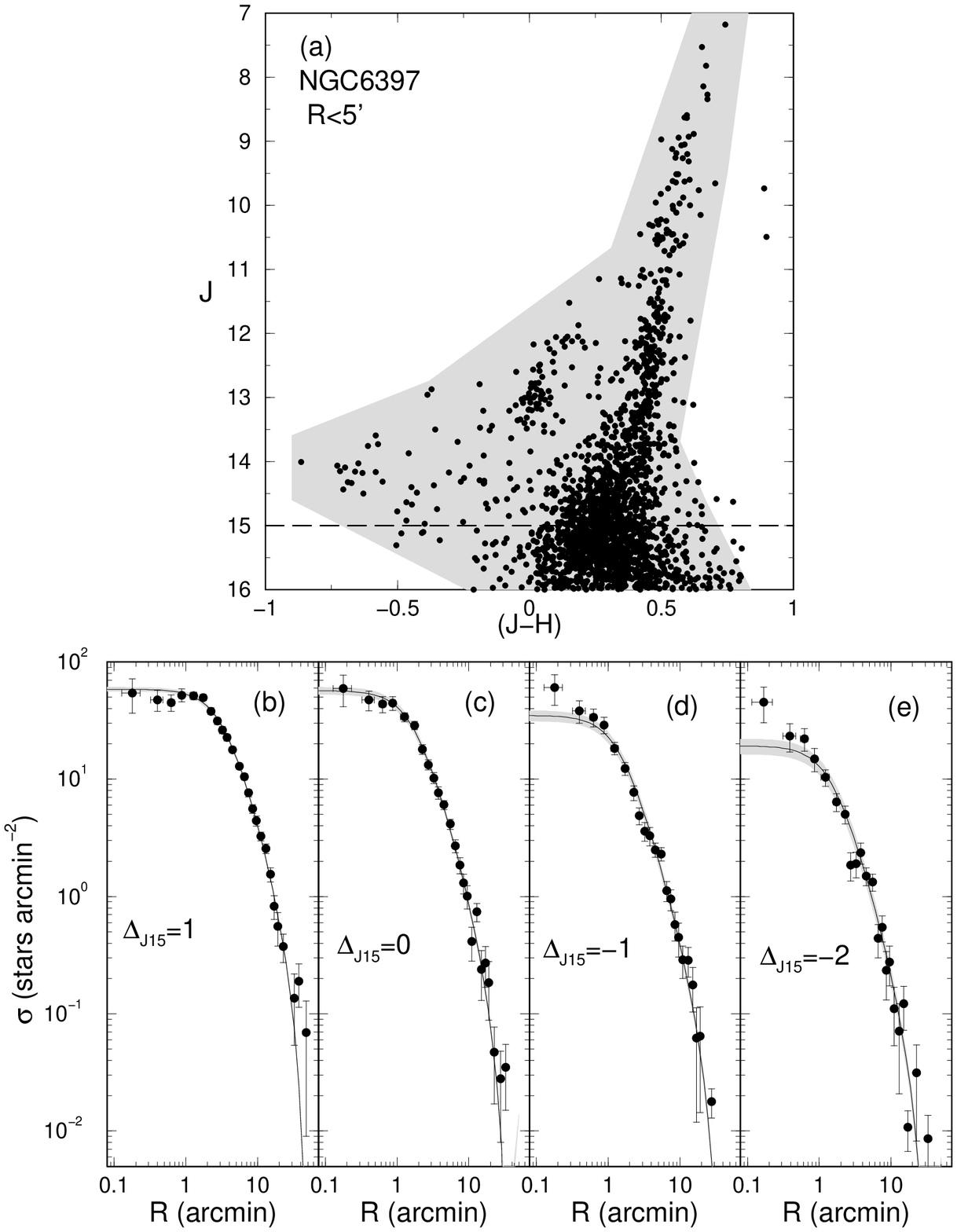}}
\caption{Structural analysis of NGC\,6397. Panel (a): decontaminated CMD of a central
($R<5\arcmin$) region. The reference magnitude $\jj=15$ is indicated by the dashed-line.
Shaded region: colour-magnitude filter. Background-subtracted RDPs for stars brighter
than $\jj<15+\Delta_{J15}$, with $\Delta_{J15}=1,~0,~-1,~-2$ are shown in panels (b) to
(d), respectively. The respective King-like fits (solid-line) together with the fit
uncertainty (shaded region) are shown.}
\label{fig8}
\end{figure}

For a better definition of the cluster sequences we apply the statistical decontamination algorithm
described in \cite{BB07}, which takes into account the relative number-densities of candidate cluster
and field stars in small cubic CMD cells with axes corresponding to the magnitude \jj\ and the colours
\jh\ and \jk. Basically, the algorithm {\em (i)} divides the full range of magnitude and colours of the
CMD into a 3D grid, {\em (ii)} computes the expected number-density of field stars in each cell based on
the number of comparison field stars with magnitude and colours compatible with those in the cell, and
{\em (iii)} subtracts the expected number of field stars from each cell. Typical cell dimensions are
$\Delta\jj=0.5$, and $\Delta\jh=\Delta\jk=0.25$, which are large enough to allow sufficient star-count
statistics in individual cells and small enough to preserve the morphology of the CMD evolutionary
sequences. The comparison field is the region located between $50\leq R(\arcmin)\leq70$, which is
beyond the tidal radius.

Field-decontaminated CMDs allow for a better definition of colour-magnitude filters, useful to remove 
stars (and artifacts) with colours compatible with those of the field which, in turn, improves the 
cluster/background contrast in RDPs and SBPs. They are wide enough to accommodate cluster MS and evolved 
star colour distributions and dynamical evolution-related effects, such as enhanced fractions of binaries 
and other multiple systems (e.g. \citealt{BB07}; \citealt{N188}).

\begin{figure}
\resizebox{\hsize}{!}{\includegraphics{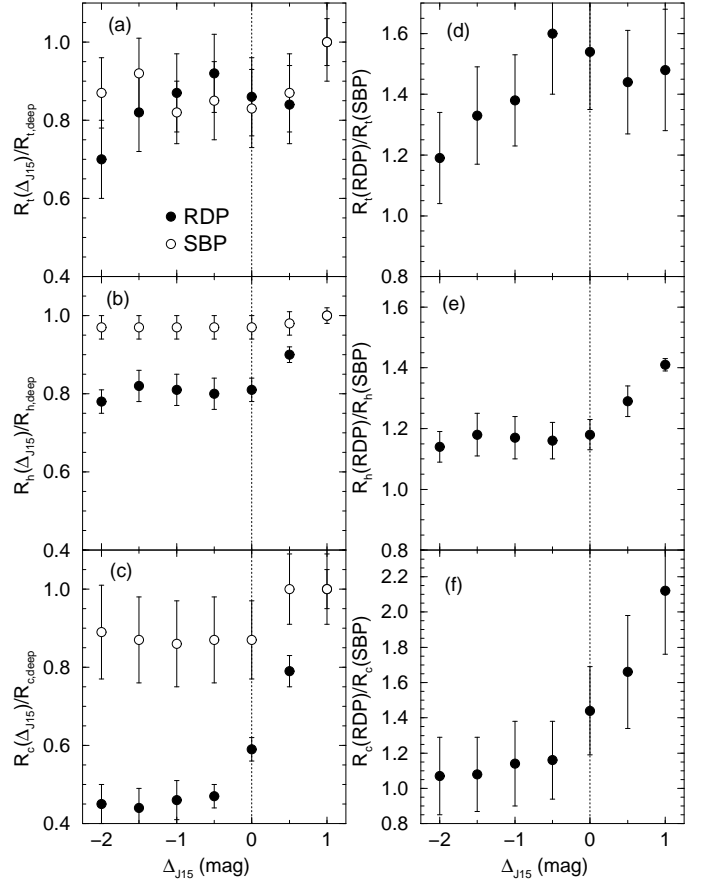}}
\caption{Left panels: RDP and SBP structural radii of NGC\,6397 as a function of $\Delta_{J15}$, 
normalised to the values measured in the deepest profile. Right panels: RDP to SBP radii ratios
(similar to Fig.~\ref{fig5}).}
\label{fig9}
\end{figure}

Figure~\ref{fig8} (panel a) displays the decontaminated CMD of a central region of NGC\,6397, with
$R<5\arcmin$, somewhat larger than the half-light radius (Table~\ref{tab3}). We take $\jj=15$ as 
reference to extract the depth-variable profiles. RDPs and SBPs are built with colour-magnitude 
filtered photometry, with the faint end varying in steps of $\Delta_{J15}=0.5$, with the deepest 
(i.e. at the available 2MASS depth) profile beginning at $\jj=16$ and the brightest one ending 
near the giant clump at $\jj=13$. The extracted profiles are fitted with the King-like function 
discussed in (Sect.~\ref{ModelSCs}). A selection of depth-limited RDPs, together with the respective 
fits and uncertainties, is shown in Fig.~\ref{fig8}, and the corresponding RDP and SBP (\jj\ band) 
radii are given in Table~\ref{tab3}. Within uncertainties, the present value of the core radius (for 
the deepest profile), $\rc(\jj)=1.4\arcmin\pm0.3\arcmin$, agrees with that derived by \citet{Cohen07}, 
using the same fit function. The near-infrared half-light radius, on the other hand, is larger than 
the optical one (H03), $\rhL(\jj)\approx1.5\rhL(V)$.

Effects of the varying magnitude depth on the radii of NGC\,6397 are examined in Fig.~\ref{fig9}.
Qualitatively, the resulting curves agree, within uncertainties, with the behaviour predicted by 
the GC models (Figs.~\ref{fig3} and \ref{fig5}). Compared to the values measured in the deepest
RDP, the tidal (panel a), half-star counts (b) and core (c) radii decrease for shallower profiles, 
especially for $\Delta_{J15}\geq-0.5$, remaining almost uniform for $\Delta_{J15}<-0.5$. In 
particular, the core radius measured in shallow RDPs (containing essentially giants) drops to 
$\approx45\%$ of its deepest value (which includes stars at the top of the MS). Consistently with
the GC models containing a spatially variable MF (Sect.~\ref{Struc}), the varying depth affects
the tidal, half and core radii, with increasing intensity. SBP radii, on the other hand, remain
essentially uniform with variable depth, consistent with the GC models (Sect.~\ref{Struc}). The
same conclusions apply to the RDP to SBP radii ratio (right panels).

\begin{table}
\caption[]{Radii of NGC\,6397 from RDPs and 2MASS SBPs}
\label{tab3}
%\tiny
\renewcommand{\tabcolsep}{0.9mm}
\renewcommand{\arraystretch}{1.25}
\begin{tabular}{cccccccc}
\hline\hline
&\multicolumn{3}{c}{RDP}&&\multicolumn{3}{c}{SBP (\jj\ band)}\\
\cline{2-4}\cline{6-8}
$\Delta_{J15}$&\rc&\rhSC&\rt&&\rc&\rhL&\rt \\
(mag)&(\arcmin)&(\arcmin)&(\arcmin)&&(\arcmin)&(\arcmin)&(\arcmin)\\
(1)&(2)&(3)&(4)&&(5)&(6)&(7)\\
\hline
$-2.0$&$1.3\pm0.1$&$3.8\pm0.1$&$33\pm5$&&$1.2\pm0.3$&$3.4\pm0.1$&$28\pm5$ \\
$-1.5$&$1.3\pm0.1$&$4.0\pm0.2$&$39\pm8$&&$1.2\pm0.3$&$3.4\pm0.1$&$30\pm5$ \\
$-1.0$&$1.3\pm0.1$&$4.0\pm0.2$&$42\pm8$&&$1.2\pm0.3$&$3.4\pm0.1$&$26\pm5$ \\
$-0.5$&$1.4\pm0.1$&$3.9\pm0.2$&$44\pm7$&&$1.2\pm0.3$&$3.4\pm0.1$&$27\pm8$ \\
$~0.0$&$1.7\pm0.1$&$4.0\pm0.1$&$41\pm4$&&$1.2\pm0.3$&$3.4\pm0.1$&$27\pm6$ \\
$+0.5$&$2.3\pm0.1$&$4.4\pm0.1$&$40\pm4$&&$1.4\pm0.4$&$3.4\pm0.1$&$28\pm8$ \\
$+1.0$&$2.9\pm0.1$&$4.9\pm0.1$&$48\pm3$&&$1.4\pm0.3$&$3.5\pm0.1$&$32\pm2$ \\
\hline
\end{tabular}
\begin{list}{Table Notes.}
\item Core and tidal radii were derived from fits of \citet{King62} functions (Sect.~\ref{Struc})
to the respective profiles. The half-star counts and half-light radii were measured
directly on the profiles.
\end{list}
\end{table}

\section{Concluding remarks}
\label{Conclu}

In this work we simulated star clusters of different ages, structure and mass functions, assuming 
that the spatial distribution of stars follows an analytical function, similar to \citet{King62} 
profile. The mass and near-infrared luminosities of each star were assigned according to a mass 
function with a slope that may depend on distance to cluster center. They form the set of models 
from which we built number-density, mass-density and surface-brightness profiles, allowing for a 
variable photometric depth. The structural parameters core, half-light, half-mass and half-star 
count, and tidal radii, together with the concentration parameter, were measured in the resulting 
radial profiles. Next we examined relations among similar parameters measured in different profiles,
and determined how each parameter depends on photometric depth.  We point out that the results 
should be taken as upper-limits, especially for open clusters, since we have considered noise-free 
photometry and a large number of stars, which produced small statistical uncertainties.

With respect to the adopted form of the radial distribution of stars, we note that \citet{King62}
isothermal sphere, single-mass profile has been superseded by more realistic models like those of
\citet{King66}, \citet{Wilson75} and \citet{EFF87}, which have been fit mostly to the SBPs of 
Galactic and extra-Galactic GCs (Sect.~\ref{ModelSCs}).
The analytical functions associated with these models are characterised by different scale radii
(among other parameters) that are roughly related to \citet{King62} radii. Thus, it is natural to
extend the scaling with photometric depth undergone by \citet{King62} radii to the equivalent ones
in the other models.

The main results can be summarised as follows. 
\begin{itemize}
\item {\em (i)} Structural parameters derived from surface-brightness profiles are essentially
insensitive to photometric depth, except perhaps the cluster radius in very young clusters.

\item {\em (ii)} Uniform mass functions also result in structural parameters insensitive to
photometric depth.

\item {\em (iii)} Number-density and mass-density profiles built with shallow photometry result
in underestimated radii, with respect to the values obtained with deep photometry. Tidal, half-star 
count and half-mass, together with the core radii are affected with increasing intensity.

\item {\em (iv)} Because of the presence of bright stars, radii underestimation increases for 
young ages.

\item {\em (v)} For clusters older than $\sim1$\,Gyr, number-density and mass-density radii
present essentially the same values; for younger ages, RDP radii become increasingly larger
than MDP ones, especially at the deepest profiles.

\item {\em (vi)} Irrespective of age, profiles deeper than the turnoff have RDP radii
systematically larger than SBP ones, especially the core.

\item {\em (vii)} The concentration parameter also changes with photometric depth, reaching 
a maximum around the turnoff. 

\end{itemize}

Most of the above model predictions were qualitatively confirmed with radii measured
in ground-based RDPs and SBPs of the nearby GC NGC\,6397.

In principle, working with SBPs has the advantage of producing more uniform structural parameters,
since they are almost insensitive to photometric depth. However, as discussed in Sect.~\ref{intro},
SBPs usually present high levels of noise at large radii. Noise that is also present in SBPs of 
clusters projected against dense fields and/or the less populous ones.  A natural extension
of this work would be to examine radial profiles built with photometry that includes observational
uncertainties, differential absorption, metallicity gradients, binaries, and star cluster models 
with a number of stars compatible with those of open clusters.

As a consequence of the wide range of distances to the Galactic (and especially extra-Galactic)
star clusters, interstellar absorption, and intrinsic instrumental limitations, the available 
photometric data for most clusters do not sample the low-mass stars. All sky surveys like 2MASS,
usually are restricted to the giant branch, or the upper main sequence, for clusters more distant
than a few kpc. In such cases, the structural parameters have to be derived from radial profiles 
built with photometry that does not reach low-mass stars. The present work provides a quantitative
way to estimate the intrinsic (i.e. in the case of photometry including the lower main sequence)
values of structural radii of star clusters observed with depth-limited photometry.

\begin{acknowledgements}
We thank the anonymous referee for helpful suggestions.
We acknowledge partial support from the Brazilian institution CNPq .
\end{acknowledgements}

%--------------------------------- References -------------

\appendix\section{Transformation of a random number distribution into
an analytical function.}
\label{Transf}

The simulations discussed in the present work depend on the transformation of a distribution 
of numbers $p(n)$, with $n$ varying in the range $0\le n\le 1$, into the analytical function 
$f(x)$, with $x$ in the range $x_{\rm min}\leq x\leq x_{\rm max}$. What results from this is
that a random sorting of numbers $n$ would produce $x$ variables with a number-frequency given 
by the pre-defined function $f(x)$. Formally, the transformation is expressed as

\begin{equation}
\label{eqA1}
f(x)\,dx=p(n)\,dn.
\end{equation}
 
In a random distribution, all numbers have the same probability, so that we can set $p(n)=1$. 
Thus, the formal relation of $n$ to $x$ is given by

\begin{equation}
\label{eqA2}
n(x)=\int_{x_{\rm min}}^{x}f(x')\,dx'/\int_{x_{\rm min}}^{x_{\rm max}}f(x')\,dx'.
\end{equation}
 
The analytical (or numerical) inversion of the latter relation gives $x=x(n)$.

For example, consider a mass function $\phi(m)=\frac{dN}{dm}\propto m^{-(1+\chi)}$, with the mass 
varying in the range $m_i\leq m(\ms)\leq m_s$. If $\chi\neq0.0$, what results from Eq.~\ref{eqA2} 
is $$n(m)=\left(m_i^{-\chi}-m^{-\chi}\right)/\left(m_i^{-\chi}-m_s^{-\chi}\right).$$ Solving this 
for $m$ produces $$m(n)=m_s/[(1-n)(m_s/m_i)^\chi+n]^{1/\chi}.$$ For $\chi=0.0$ the result is
$m(n)=m_i\,(m_s/m_i)^n$.

In the case of the three-parameter King-like profile (Eq.~\ref{eq1} in Sect.~\ref{ModelSCs}), 
the first step in the transformation, represented by Eq.~\ref{eqA2}, results in the $\rm n=n(R)$ 
function given by Eq.~\ref{eq2} (Sect.~\ref{ModelSCs}). However, the latter relation cannot be 
analytically inverted, so that this task has to be done numerically, using the curves plotted in 
Fig.~\ref{fig1} (panel a). 

%\end{appendix}

\end{document}